\documentclass[aps,prd,twocolumn,preprintnumbers,groupedaddress,superscriptaddress,floatfix,tightenlines,reprint,nofootinbib]{revtex4-2}
\usepackage{mathrsfs,natbib}
\usepackage[dvipsnames]{xcolor}
\usepackage{verbatim}
\usepackage{bm,physics}
\usepackage{enumerate}
\usepackage{epsf,amssymb,bbm,amsbsy,amsfonts,amssymb,amsmath}
\usepackage{hyperref}
\usepackage{graphicx}
\allowdisplaybreaks

\newcommand{\be}{\begin{eqnarray}}
\newcommand{\ee}{\end{eqnarray}}
\newcommand{\beq}{\begin{equation}}
\newcommand{\eeq}{\end{equation}}

\hypersetup{
    colorlinks=true,       
    linkcolor=red,         
    citecolor=blue,        
    filecolor=magenta,     
    urlcolor=blue          
}

\begin{document}

\title{Confinement and graded partition functions for \texorpdfstring{$\mathcal{N}=4$}{N=4} SYM}
\author{Aleksey Cherman}
\email{acherman@umn.edu}
\author{Aditya Dhumuntarao} 
\email{dhumu002@umn.edu}
\affiliation{School of Physics and Astronomy, University of Minnesota
  Minneapolis, MN 55455}

\begin{abstract}
	Gauge theories with confining phases at low temperatures tend to deconfine at high temperatures.  In some cases, for example in supersymmetric theories, confinement can persist for all temperatures provided the partition function includes a grading by $(-1)^F$. When it is possible to define partition functions which smoothly interpolate between no grading and $(-1)^F$ grading, it is natural to ask if there are other choices of grading that have the same effect as $(-1)^F$ on confinement.  We explore how this works for $\mathcal{N}=4$ SYM on $S^1\times S^3$ in the large $N$ limit at both small and large coupling.  We find evidence for a continuous range of grading parameters that preserve confinement for all temperatures at large coupling, while at small coupling only a discrete set of gradings preserves confinement.
\end{abstract}

\maketitle
	\textbf{Introduction}. Gauge theories tend to deconfine at high temperatures.  
	This statement can be made precise in theories that have a center symmetry,
	such as $SU(N)$ Yang-Mills theory, see e.g.~\cite{Gross:1980br,Gaiotto:2014kfa}.  Studying the
	properties of a $4d$ theory on a spatial manifold $M_3$ at finite temperature
	is equivalent to asking about the behavior of the Euclidean path integral on
	$M_3 \times S_\beta^1$ with periodic boundary conditions for bosons and
	anti-periodic boundary conditions for fermions. This path integral gives the
	thermodynamic partition function
	\begin{align}
		Z(\beta) = \tr e^{-\beta H}
		\label{eq:thermal_partition_function}
	\end{align}
	where $\beta = 1/T$.  Wilson loops that wind around the thermal circle, $\tr
	\Omega = \mathcal{P} e^{i \int_{S^1} A}$ transform by a $\mathbb{Z}_N$ phase
	under center symmetry, $\tr \Omega  \to e^{2\pi i /N} \tr  \Omega$. In pure
	$SU(N)$ YM theory, it is known that $\langle \tr \Omega \rangle = 0$ at low
	temperatures, see e.g.~\cite{Greensite:2003bk}, but it is non-zero at high
	temperatures \cite{Gross:1980br}. This implies that the high and low
	temperature regimes are separated by a phase transition in the thermodynamic
	limit. Note however that there are actually two physically distinct ways to
	take the thermodynamic limit for $SU(N)$ YM theory.  The standard way is
	take the volume of $M_3$ to infinity. Alternatively, one can take $N \to
	\infty$ with the 't Hooft coupling $\lambda = g_\text{YM}^2 N$ and other
	parameters such as the spatial volume held fixed
	\cite{Sundborg:1999ue,Aharony:2003sx,Aharony:2005bq}. There is a
	temperature-driven deconfinement transition in both of these two
	thermodynamic limits.  In this paper we will discuss the deconfinement phase
	transition in the the large $N$ thermodynamic limit.

	The deconfinement phase transition is rather difficult to evade by, e.g.,
	varying parameters controlling the matter content.  Indeed, in the large $N$
	limit, the deconfinement transition is forced by the fact that the confining
	phase features a Hagedorn density of states.  This means that the density of
	states, $\rho(E)$, grows exponentially with the energy $\rho(E) \sim e^{+\beta_H E}$,
	where it is assumed that $E\sim N^0$ is large compared to the characteristic
	mass scale of the theory, and $\beta_H > 0$, the Hagedorn temperature,
	depends on the matter content of the theory, but can be estimated as
	$\beta_{H} \sim \textrm{min}(\Lambda^{-1}, R)$ where $\Lambda$ is the large
	scale and $R$ is the characteristic size of the spatial box. Given this density 
	of states, the thermodynamic partition function cannot be analytic
	at $\beta = \beta_H$. Thus the theory must have deconfinement transition at
	some $\beta_c \ge \beta_H$ at large $N$. 
	
	However, the deconfinement phase transition can be evaded if one considers some 
	\emph{graded} partition functions.  The most famous example is 
	\begin{align}
		\tilde{Z}(\beta) = \tr (-1)^F e^{-\beta H} \,
		\label{eq:graded_partition_function}
	\end{align}
	where $F$ is fermion parity.  This partition function is calculated by a
	path integral with periodic boundary conditions for all fields on
	$S^1_\beta$.  If the distribution of bosonic and fermionic states are
	uncorrelated, a $(-1)^F$-graded partition function will also exhibit phase
	transitions as a function of $\beta$ in the thermodynamic limit. However, if
	the spectra of bosonic and fermionic states are related in a sufficiently
	precise way, there will be large cancellations and the graded density of
	states $\tilde{\rho}(E)$ can have sub-exponential growth at large $E$ at arbitrary $\beta$. Then
	$\tilde{Z}(\beta)$ can avoid the Hagedorn instability and the deconfinement
	phase transition.
	
	This naturally occurs in supersymmetric gauge theories, 
	where the bosonic and fermionic states with energies $E>0$ necessarily come 
	in degenerate multiplets in the infinite-volume limit. Surprisingly, deconfinement 
	phase transitions of \eqref{eq:graded_partition_function} 
	can sometimes be avoided even without supersymmetry. Indeed,
	this is the case for adjoint QCD, YM theory coupled to $N_f>1$ adjoint
	fermions \cite{Kovtun:2007py,Basar:2013sza,Basar:2014jua,Cherman:2018mya,Cherman:2019ecx}, 
	and even in some theories with fundamental fermions, 
	which require more complicated gradings \cite{Kanazawa:2019tnf}. In such theories, 
	$\tilde{Z}(\beta)$ is smooth, and hence the theory remains 
	confining at all temperatures.

	In this paper, we want to study what happens to deconfinement transition in
	theories where it is possible to smoothly interpolate between
	\eqref{eq:thermal_partition_function} and
	\eqref{eq:graded_partition_function}, both at large and small 't Hooft
	coupling.  This is most easily done for the large $N$ limit of $4d$
	$\mathcal{N}=4$ super Yang-Mills theory, formulated in Euclidean signature
	on $S_R^3 \times S^1_{\beta}$, where $R$ is the radius of $S^3$. Indeed, it
	is known that without any grading, large $N$, $\mathcal{N}=4$ SYM has a
	deconfinement as a function of $\beta/R$, and the critical value of
	$\beta/R$ is known both at small
	\cite{Sundborg:1999ue,Aharony:2003sx,Aharony:2005bq} and
	large \cite{Maldacena:1997re,Gubser:1998bc,Witten:1998qj,Witten:1998zw} 't
	Hooft coupling\footnote{More generally, the critical value of $\beta/R$ is
	bounded from below by the Hagedorn temperature, which is known for all
	$\lambda$ \cite{Harmark:2017yrv,Harmark:2018red}.} due to the AdS/CFT
	correspondence. Supersymmetry suggests that the theory should be in a
	confined phase for all $\beta/R$ if the partition function includes a
	grading by $(-1)^F$, and indeed this has been verified at large coupling
	in Ref.~\cite{Witten:1998zw}. Finally, $\mathcal{N}=4$ SYM
	has a continuous internal global symmetry, $SO(6)_R$, which, as we will show,
	can be used to define a grading that smoothly interpolates between
	\eqref{eq:thermal_partition_function} and
	\eqref{eq:graded_partition_function}.  We define this grading below and
	study the resulting phase structure at small and large 't Hooft coupling.

	{\bf Graded partition functions. } In a conformal field theory (such as
	$\mathcal{N}=4$ SYM) on $S^3_R \times S^1_{\beta}$, the partition function
	can be viewed as $Z(\beta/R) = \sum_{\Delta} e^{-\Delta \beta/R}$, where the
	sum runs over the scaling dimensions of local operators, so that the
	Hamiltonian is related to the dilation operator $H = D/R$.  In theories with
	global symmetries it is also possible to turn on chemical potentials $\mu_i$
	associated to conserved charges $Q_i$, yielding
	\begin{align}
		Z(\beta, \mu_i) = \tr e^{\beta \sum_{i} \mu_i Q_i} e^{-\beta  H} \,.
	\end{align}  
	In the specific case of $\mathcal{N}=4$ SYM, there is an $SO(6)$ R-symmetry,
	and so one can turn on three independent chemical potentials associated with
	the three Cartan generators of $SO(6)_R$. We can consider $\mathcal{N}=4$ SYM
	as $\mathcal{N}=1$ SYM coupled to three adjoint chiral supermultiplets, then
	the three complex scalar fields $\phi_1, \phi_2, \phi_3$ in $\mathcal{N}=4$ SYM
	are the lowest components of the three chiral supermultiplets. The three
	scalars transform in the $\mathbf{6}$ representation of $SO(6)_R$, and
	following the conventions of Ref.~\cite{Yamada:2006rx}, we assemble the
	three scalars into a six-dimensional vector
	$(\phi_1,\phi_1^{*},\phi_2,\phi_2^{*},\phi_3,\phi_3^{*})^T$, and write the
	Cartan generators in the $\mathbf{6}$ representation as
	\begin{align}
		Q^{\mathbf{6}}_1 &= \textrm{diag}(+1,-1,0,0,0,0) \\
		Q^{\mathbf{6}}_2 &=  \textrm{diag}(0,0,+1,-1,0,0) \\
		Q^{\mathbf{6}}_2 &= \textrm{diag}(0,0,0,0,+1,-1) \, ,
	\end{align}
	Then $2Q^{\mathbf{6}}_i$ acts with charge $2$ on $\phi_i$, and the
	superconformal R-charge $r$ can be written as $r =
	\frac{2}{3}(Q_1+Q_2+Q_3)$, see e.g.~\cite{Benini:2018ywd}.  We will set
	$\mu_1 = \mu_2 = \mu_3  = 2 \mathrm{i} \theta/\beta$ and study the partition function
	\begin{align}
		Z(\beta,\theta) &= \tr e^{2\mathrm{i}\theta \sum_{i=1}^{3} Q_i} e^{-\beta H}
		= \tr e^{3 \mathrm{i} \theta r} e^{-\beta H} \,.
		\label{eq:interpolating_partition_function}
	\end{align}
	This graded partition function is periodic in $\theta$ with period $2\pi$.
	It implements a simple interpolation from
	\eqref{eq:thermal_partition_function} at $\theta = 0$ to
	\eqref{eq:graded_partition_function} at $\theta = \pi$, thanks to the fact
	that in $\mathcal{N}=4$ SYM there is a spin-charge relation $e^{3\pi\mathrm{i} r} =
	(-1)^F$, see e.g.~\cite{Copetti:2020dil}.  Charge conjugation symmetry further relates $r$ to $-r$, so that the $\theta$ peridicity is reduced to $\theta \simeq \theta +\pi$.

	\begin{figure}[t]
		\includegraphics[width=.45\textwidth]{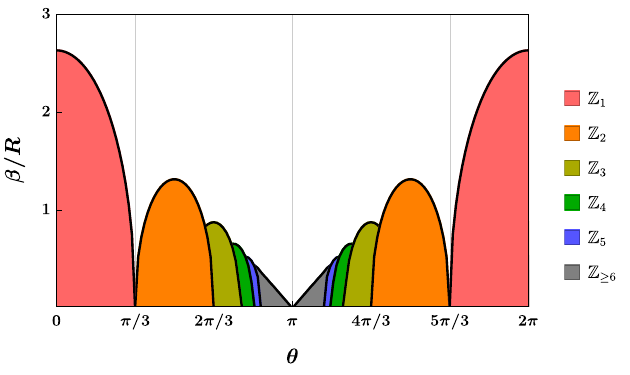}
		\caption{ %
		Large $N$ phase diagram of $\mathcal{N}=4$ SYM theory on $S_R^3$
		for $\lambda \to 0$ as a function of inverse temperature
		$\beta$ and grading $\theta$.  The theory is in a confined phase in
		the white region, and is partially confined in the
		colored regions.  In between $\theta = \pi/3$ and $\theta = 5\pi/3$
		center symmetry is only partially broken, with the unbroken subgroup of
		$\mathbb{Z}_N$ indicated by the color scheme in the legend. The theory
		confines at all $\beta/R$ when $\theta = \pi/3,\pi,5\pi/6$.}
		\label{fig:theta}
	\end{figure}

	{\bf Zero coupling.}  Let us determine the $\beta$ and $\theta$ dependence
	of $Z$ in \eqref{eq:interpolating_partition_function} in the zero coupling
	limit $\lambda \to 0$, at large $N$. The partition function is
	given by a matrix integral which can be interpreted as an integral over the
	holonomy $\Omega$ \cite{Aharony:2003sx}, which implements the Gauss law constraint
	\cite{Gross:1980br}. The matrix
	integral can be reduced to an integral over the eigenvalues of the holonomy
	\begin{align}
	Z(\beta/R, \{\mu_i\}) = \int \left(\prod_{k} d \lambda_k \right)
	 e^{-\mathcal{S}_{\rm eff}(\lambda_k; \frac{\beta}{R},\{\mu_i\})}
	\end{align} 
	where $\lambda_k$ are the eigenvalues, $\{\mu_i\}$ are the three independent
	chemical potentials for mutually commuting charges in $SO(6)_R$, and
	$\mathcal{S}_{\rm eff}$ is an effective action.  The holonomy eigenvalues
	take values on a circle, and the effective action can be written in terms of
	Fourier coefficients $\rho_n$ of the eigenvalue distribution $\rho$
	\begin{equation}
		\mathcal{S}_{\text{eff}} = N_c^2 \sum_{n=1}^{\infty} V_n \rho_n^2
	\end{equation}
	 where $\rho_{n} \equiv
	\int_{-\pi}^\pi\dd\alpha\rho(\alpha)\cos(n\alpha)$ and the $V_n$ coefficients
	are given by
	\begin{align} 
		V_n \equiv \frac{1}{n} &\left(1- z_B(x^n,\{\mu_i\}) + (-1)^{n}z_F(x^n,\{ \mu_i\})]  \right).
	\end{align}
	The parameter $x = e^{-\beta/R}$, and the functions $z_B = z_V +z_S, z_F$ are
	related to the single-particle partition functions for the massless
	vector, scalar, and fermion fields of $\mathcal{N}=4$ SYM on $S^3$ \cite{Yamada:2006rx}:
	\begin{align}
		z_{V}(x)&=\frac{6x^2-2x^3}{(1-x)^3}\,,\\
		z_{S}(x,\{\mu_i\})&=\frac{x+x^2}{(1-x)^3}\sum_{a=1}^3(x^{R\mu_i}+x^{-R\mu_i})\,,\\
		z_{F}(x,\{\mu_i\})&=\frac{2 x^{3/2}}{(1-x)^3}\prod_{a=1}^{3}(x^{\frac{1}{2}R\mu_i}+x^{-\frac{1}{2}R\mu_i})\,.
	\end{align}
	We now set  $\mu_i = \frac{2 \mathrm{i} \theta}{\beta}$, so that
	\begin{subequations}
		\begin{align}
			z_S &= 6 \frac{x+x^2}{(1-x)^3}\cdot\qty[\cos(2\theta)]\, , \\
			z_{F} &=16 \frac{ x^{3/2}}{(1-x)^3}\cdot\qty[\cos(\theta)]^3\,.
	\end{align}
	\label{eq:twisted_letters}
\end{subequations}
	
	So long as all of the coefficients $V_n$ are positive, the matrix integral
	is dominated by the center-symmetric $\rho_n = 0$ minimum. When there is
	value of $n$ for which $V_n$ becomes negative there is a phase transition.
	We can interpret these phase transitions as center-breaking phase
	transitions where $\mathbb{Z}_N$ is spontaneously broken to an (approximate)
	subgroup $\mathbb{Z}_{n}$.  (For an early discussion of partial deconfinement
	see e.g.~\cite{Myers:2007vc}.)
	For large $\beta/R$, corresponding to $x \ll 1$, it is easy to verify that
	$V_n > 0$ for all $n$, but for small $\beta/R$ some $V_n$ generically become
	negative.  Figure~\ref{fig:theta} is a plot of the phase diagram as a function of
	$\theta$ and $\beta/R$. Center symmetry is preserved above the black curve,
	and is broken either completely or to a subgroup of $\mathbb{Z}_N$ below the
	black curve.  
	
	There are two interesting features in Fig.~\ref{fig:theta}.  First, the lack
	of deconfinement for any $\beta$ at $\theta = \pi$ is consistent with our
	original expectations:  it amounts to working with a $(-1)^F$ graded
	partition function. However, the theory also remains confining for all
	$\beta/R$ if $\theta = \pi/3, 5\pi/3$.  In the region
	$(\pi/3,\pi)\cup(\pi,5\pi/3)$ the system is in a (partially) center-broken
	phase for sufficiently small values of $\beta/R$. The dependence of the
	deconfinement temperature on $\theta$ is highly non-monotonic.

	What should we expect as $\lambda$ is increased from zero?  At a heuristic
	level, increasing the coupling $\lambda$ should increase the fluctuations of
	the Polyakov loop eigenvalues, and it is natural to expect such fluctuations
	to increase the range of $\beta/R$ values where all traces of powers of the
	Polyakov loop vanish.  It would be interesting but challenging to directly
	compute corrections to our results in powers of $\lambda$, see e.g.
	\cite{Aharony:2005bq,Mussel:2009uw} for the challenges that arise already in
	pure YM theory.  However, one of the special features of $\mathcal{N}=4$ SYM
	is that the $\lambda \to \infty, N \to \infty$ limit is just as calculable
	as the $\lambda \to 0, N\to \infty$ limit thanks to the AdS/CFT
	correspondence
	\cite{Maldacena:1997re,Gubser:1998bc,Witten:1998qj,Witten:1998zw}. In what
	follows, we take advantage of AdS/CFT and study the $\theta$ dependence of
	the phase diagram at large 't Hooft coupling.

	{\bf Infinite coupling.} 
	At large $N$ and $\lambda$, $\mathcal{N}=4$ $SU(N)$ SYM is believed to
	have a dual gravitational description
	\cite{Maldacena:1997re,Gubser:1998bc,Witten:1998qj,Witten:1998zw}. The
	dependence of the theory on R-charge chemical potentials has been
	extensively explored in the literature, see e.g.
	\cite{Chamblin:1999tk,Chamblin:1999hg,Witten:1998qj,Witten:1998zw,Yamada:2006rx}.
	The AdS/CFT dictionary for conserved charges implies that to study
	$\mathcal{N}=4$ SYM with chemical potentials for $R$ symmetry, we should
	consider the truncation of Type IIB supergravity on AdS$_5 \times S^5$ to a
	$5d$ Einstein gravity theory coupled to three $U(1)$ gauge fields $(A_{I})$, $I
	= 1,2,3$, associated with the $U(1)^3$ Cartan subgroup of $SO(6)_R$
	\cite{Chong:2005da,Chong:2006zx,Chong:2005hr,Cvetic:1999ne,Cvetic:2004ny}:
	\begin{align}
		\mathcal{I} = - \frac{1}{2\kappa^2}&\int \dd^{5}x \sqrt{-g}\left[{R} +
		\frac{12}{\ell^2} - \frac{\ell^2}{4}\sum_{I=1}^3 (F_I)^2\right]\label{eqn:gaugedSUGRA}
	\end{align}
	where $(F_I)_{MN}=\partial_{M} (A_I)_N - \partial_N (A_I)_M$ are the field strengths of the
	three gauge fields, $M,N =
	1,\ldots, 5$, $\kappa$ is related to the $5d$ Newton constant $G_5$ via
	$\kappa=8\pi G_5$, and the cosmological constant is $\Lambda =
	-6/\ell^2$. 

	The boundary values of bulk gauge fields act as sources for conserved
	currents in the boundary theory, and our goal is to turn on equal chemical
	potentials for each of the three Cartan charges.  For this purpose we can
	simply set $A^I=A$ above, and then the bulk action reduces to an Einstein-Maxwell-AdS$_5$ theory
	\begin{align}\label{eq:SUGAR}
		\!\mathcal{I} = -\frac{1}{2\kappa^2}\int \dd^5 x \sqrt{-g} &\left[{R} + \frac{12}{\ell^2} 
		- \frac{3\ell^2}{4} F_{MN}F^{MN}\right].
	\end{align}

	The Einstein-Maxwell-AdS$_5$ system has been extensively explored in
	\cite{Chamblin:1999tk,Chamblin:1999hg}, and has
	 Reissner-Nordstrom-AdS${}_5$ black hole solutions\footnote{Our normalization convention
	is different from \cite{Chamblin:1999tk,Chamblin:1999hg}, so that
	$A_\text{there} = {\sqrt{4/3\ell^2}} A_\text{here}.$} with parameters $(L_m,L_q)$ 
	which are related to the ADM mass and the charge density respectively. 
	The charged black hole solution reads
	\begin{align}
		\dd s^2 &= - f(r)\dd t^2 + \frac{\dd r^2}{f(r)}+r^2\dd\Omega_3^2\label{eqn:metric},\\
		f(r) &= 1 + \frac{r^2}{\ell^2} - \frac{L^2_m}{r^{2}} + \frac{L_q^4}{r^{4}}\label{eqn:blackening},\\
		A_t(r)&=\qty(\mu - \frac{L_q^2}{\ell r^2}) \,.\label{eqn:LTgaugefield}
	\end{align}
	This non-supersymmetric family of non-rotating solutions is expected to
	correspond to spatially-homogeneous equilibrium states of the dual field
	theory.  

	We briefly review how the thermodynamic parameters of the bulk solution,
	i.e., the inverse temperature and chemical potential, map to the
	corresponding field theory parameters.  First, we recall that $\ell$ can be
	identified with the radius of the $S^3$ spatial manifold in the dual field
	theory \cite{Yamada:2006rx}.  To identify $\beta$, we pass to Euclidean signature $t\to
	\mathrm{i}\tau$, and recall that the metric is free of conical singularities
	provided $\tau$ is periodic with period
	\begin{equation}
		\beta = \frac{4\pi}{\partial_r f(r)}\eval_{r_+} = 
		\frac{2\pi \ell^2 r_+}{\ell^2(1-{(L_q/r_+)}^4)+2 r_+^2}\label{eqn:HawkingPer}
	\end{equation}
	where $r_+$ is the outer horizon corresponding to the largest positive root
	 	of $f(r_+)=0$.  The Hawking temperature $T = 1/\beta$ is then identified
	 	with the temperature of the dual field theory. Next, the chemical
	 	potential is determined by the asymptotic value of the field strength
	 	along the $\tau-r$ disk, $D_2$: 
	\begin{align}
		\int_{D_2} F = \int_0^\beta \dd \tau \int_{r_+}^\infty \dd r \,F_{r\tau}\,= \mathrm{i}\beta\mu. \label{eqn:fugacity}
	\end{align}
	The gauge field must be regular at $r_+$, meaning that $A_\tau(r_+)=0$, which yields the identification
	\begin{equation}
		\mu\,=\frac{L_q^2}{\ell r_+^2}.\label{eqn:gravitychempot}
	\end{equation}

The confinement/deconfinement phase transition is mapped onto a gravitational
	phase transition in the following way. The solution with $L_m=L_q = 0$
	should be thought of as thermal AdS$_5$, and is dual to the confining phase
	of the dual gauge theory, whereas solutions with $L_m^2 \in \mathbb{R}_{+}$
	are dual to the deconfined phase \cite{Witten:1998zw}.  The difference of
	the free energies of these two phases is
	\cite{Chamblin:1999tk,Chamblin:1999hg,Emparan:1999pm}
	\begin{align}
		\mathcal{F}_{\rm deconfined} - \mathcal{F}_{\rm confined}  
		=\frac{\pi r_+^2}{8 G_5}\qty(1-(\ell\mu)^2 - \frac{r^2_+}{\ell^2} ).
		\label{eq:large_free_energy}
	\end{align}
	When this quantity changes sign, there is a deconfinement phase transition
	from the point of view of the field theory \cite{Witten:1998qj,Witten:1998zw}, 
	which is realized as a Hawking-page phase
	transition in the gravitational theory \cite{Hawking:1982dh}.  
	
	To map out the dependence of the phase transition temperature on the
	parameter $\theta$ in \eqref{eq:interpolating_partition_function}, we study
	how \eqref{eq:large_free_energy} behaves as a function of an appropriate
	imaginary chemical potential.  We take $\mu \to \mathrm{i}\tilde{\mu} =
	\frac{2\mathrm{i}\theta}{\beta}$ while keeping $r_+$, which requires sending
	$L_q^2 \to \mathrm{i} \tilde{L}_q^2$.  Then $A_{\tau}$ is real and continues
	to satisfy $A_{\tau}(r_+) = 0$.  The on-shell Euclidean
	action becomes
	\begin{equation}
	\!\!\!	\mathcal{F}_{\rm deconfined} - \mathcal{F}_{\rm confined} 
		 = \frac{\pi r_{+}^2}{8 G_5}\qty(1+\qty(\frac{2\ell\theta}{\beta})^2 - \frac{r^2_+}{\ell^2} ).
	\end{equation}
	This expression is manifestly invariant under $\theta \to -\theta$. In
	addition, we should understand the $\theta$ dependence in this formula and
	the ones that follow mod $2\pi$.  In field theory, this basic fact follows
	from charge quantization. To see how this is matched in the bulk,  we recall
	that Type IIB string theory has an NS-NS two-form gauge field $B$.  One can
	turn on a flat $B$ potential (that is, obeying $d B = 0$) in the solutions
	discussed above at no cost in energy, with an arbitrary value of the $2\pi$
	periodic parameter $\alpha$
	\cite{Witten:1998zw,Aharony:1998qu,Aarts:2010ky}:
	\begin{align}
		\alpha = \int_{D_2} B.
	\end{align}
	The periodicity of the chemical potential under shifts of $2\pi \mathrm{i}/\beta$ is ensured
	by the shift freedom of $\alpha\to\alpha + 2\pi$.

	\begin{figure}[t]
		\includegraphics[width=.45\textwidth]{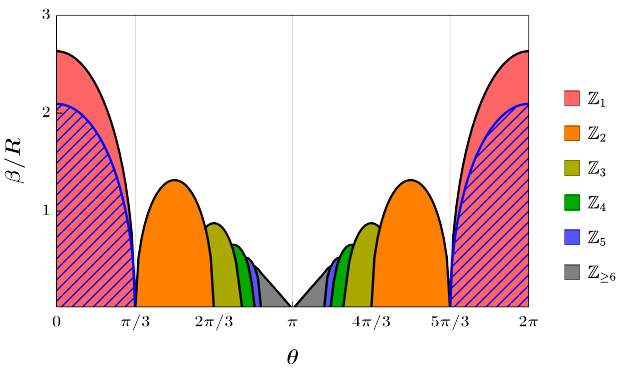}
		\caption{ %
		Large $N$ phase diagram of $\mathcal{N}=4$ SYM theory on
		 $S_R^3$ as a function of inverse temperature $\beta$
		and grading $\theta$ at zero and infinite 't Hooft coupling
		$\lambda$. The grading $\theta$ interpolates between the thermal and
		$(-1)^F$ graded partition functionsf. At $\lambda = 0$ the theory has a
		deconfinement phase transition along the black curve, and has at least
		partially broken center symmetry everywhere in the colored regions.  The
		analogous phase transition curve due to known bulk solutions at large
		coupling $\lambda \to \infty$ is shown in blue, with center symmetry
		broken everywhere in the hatched region.  Note that the deconfined
		region at large-coupling is a subset of the deconfined
		region at zero coupling.}
		\label{fig:small_large_plot}
	\end{figure}

	It is now straightforward to determine the phase transition temperature as a function of $\theta$.
	The critical temperature, $T_c$, is determined by the solution of $\mathcal{F}_{\rm deconfined} -
	\mathcal{F}_{\rm confined} = 0$, which reduces to  
	\begin{equation}\label{eqn:HP-twisted}
		1 = \frac{\ell^2}{r_+^2}\qty[1+\left(\frac{2\ell\theta}{\beta_c}\right)^2].
	\end{equation}
	Using \eqref{eqn:HawkingPer}, we may trade $r_{+}$ for $\beta_c$, replace $\ell$ by $R$, and find the transition temperature in field theory variables
	\begin{equation}\label{eqn:HP-beta}
		{\beta_c}(\theta) = \frac{2\pi R}{3}\sqrt{1 - \qty(\frac{\theta}{\pi/3})^2}.
	\end{equation}
	When $\theta = 0$, corresponding to conventional thermal field theory we
	recover the standard Hawking-Page phase transition for uncharged SAdS$_5$
	black holes $\beta_c = 2\pi R/3$ \cite{Hawking:1982dh}. As $\theta$ is
	increased the deconfinement temperature decreases.  At $\theta = \pi/3 
	\textrm{ mod } 2\pi$ the critical
	temperature becomes infinite, and there are no sensible
	Reissner-Nordstrom-AdS$_5$ black hole solutions because $\beta_c$ becomes
	complex.   The fact that $\mathcal{N}=4$ SYM is confined for all $\beta$
	when $\theta = \pi$, corresponding to $(-1)^F$ graded partition function,
	has been known since Ref.~\cite{Witten:1998zw}. But our analysis here suggests that there
	is actually a whole window of $\theta$ values, namely 
	\begin{equation}
		\theta  \in \left[ \frac{\pi}{3},\, \frac{5\pi}{3} \right] 
	\end{equation}
	where  $\mathcal{N}=4$ $SU(N)$ SYM apparently remains confined
	for all values of $\beta/R$ at large $\lambda$. We compare the zero and
	large coupling phase diagrams in Fig.~\ref{fig:small_large_plot}.

	We should emphasize that to reach \eqref{eqn:HP-beta} we have only studied
	the known (analytically-continued) solutions that correspond to homogeneous
	equilibrium states in the dual field theory.  It would be very interesting to
	see whether there might be some sort of previously unknown (perhaps
	multi-center) black objects which would be associated to the
	partially-confined phases\footnote{There is another
	notion of `partial deconfinement/sub-matrix deconfinement' recently
	discussed in Refs.~\cite{Hanada:2018zxn,Berenstein:2018lrm,Hanada:2019czd,Hanada:2019kue,Hanada:2019rzv,Hanada:2020uvt,Watanabe:2020ufk}.} we saw at small coupling,
	see e.g. \cite{ArabiArdehali:2019orz} for a discussion in the context of the
	superconformal index.

	{\bf Conclusions.} In $\mathcal{N}=4$ SYM theory, it is possible to smoothly
	interpolate from the standard thermodynamic partition function to a
	$(-1)^F$-graded partition function by taking advantage of the $SO(6)$
	R-symmetry.  We studied the simplest such generalized partition function,
	where the grading factor is $e^{3 \mathrm{i} \theta\,  r}$ uses the
	superconformal R-charge $r$.  We found that the large $N$ theory is in
	the confined phase on $S^3_R \times S^1_{\beta}$ when $\theta =
	\pi/3,\pi,5\pi/3$ both at small and large coupling. The result for $e^{3\pi
	\mathrm{i} r} = (-1)^F$ was already known. The other two points with
	complete confinement for all temperatures at both small and large 't Hooft
	coupling correspond to the more complicated  grading $(-1)^F e^{\pm 2\pi
	\mathrm{i} r}$. The small coupling and large coupling corners of parameter
	space differ in the behavior for $\theta \in (\pi/3,\pi)\cup(\pi,5\pi/3)$.
	At small coupling, the phase transition temperature depends
	non-monotonically on $\theta$ in the region $\theta \in
	(\pi/3,\pi)\cup(\pi,5\pi/3)$, with center symmetry partially broken for high
	enough temperatures, with the critical temperature diverging as $\theta \to
	\pi$.  At large coupling, on the other hand, we found no evidence of such
	rich behavior:  the theory appears to be in the confined phase for all
	temperatures for $\theta \in [\pi/3, 5\pi/3]$.
	
	It is interesting to connect our results to some recent studies of
	deconfinement in the superconformal index of $\mathcal{N}=4$ SYM.  At the
	outset, however, we should stress that our graded partition function is
	quite different from the superconformal index of $\mathcal{N}=4$ SYM
	\cite{Kinney:2005ej}, see e.g. Refs.~\cite{Rastelli:2016tbz,Gadde:2020yah} for recent reviews. The partition function
	we focus on here receives contributions from all of the states in the theory
	and depends non-trivially on the 't Hooft coupling.  On the other hand,  the
	superconformal index is a special $S^3_R \times S^1_{\beta}$ partition
	function which is designed to only receive contributions from $1/4$ BPS
	states (in $\mathcal{N}=1$ language). It is independent of the 't Hooft
	coupling, so that its exact form for all values of $\lambda$ and $N$ can be
	determined in terms of a matrix integral.  The superconformal index involves
	two chemical potentials (and corresponding fugacities $p,q$) built out of
	appropriate combinations of $\Delta$ R-charge $r$ and the left and right
	moving angular momenta $j_1,j_2$. When the fugacities are real, the
	superconformal index $\mathcal{I}$ is always in the `confining phase',
	meaning that $\log \mathcal{I} \sim \mathcal{O}(1)$ \cite{Kinney:2005ej}. It
	was recently discovered that $\log \mathcal{I}$ can become
	$\mathcal{O}(N^2)$ if  the fugacities are rotated into the complex plane, so
	that the superconformal index can be used to study deconfinement and black
	holes, see Ref.~\cite{Hosseini:2017mds,Cabo-Bizet:2018ehj,
	Choi:2018hmj,Choi:2018vbz,Closset:2017zgf,Benini:2018mlo,Benini:2018ywd,
	Lezcano:2019pae,Lanir:2019abx,Honda:2019cio,ArabiArdehali:2019tdm,
	Cabo-Bizet:2019osg,Amariti:2019mgp,Cabo-Bizet:2019eaf,ArabiArdehali:2019orz,
	Cabo-Bizet:2020nkr,Agarwal:2020zwm,Benini:2020gjh,GonzalezLezcano:2020yeb,
	Zaffaroni:2019dhb,Copetti:2020dil,Goldstein:2020yvj,Cabo-Bizet:2020ewf}.
	While the superconformal index and our partition functions are distinct,
	they do share an interesting parallel:  the grading $(-1)^F e^{\pm 2\pi i
	r}$ plays the same role in both cases as being on the boundary between
	deconfinement for some $\beta/R$ and confinement for all $\beta/R$.  This
	was recently shown in Ref.~\cite{Copetti:2020dil}, which studied the
	superconformal index as a function of $p = q = y e^{i\psi}$, finding a
	deconfined region for $\psi \in (-4\pi,-3\pi) \cup  (-3\pi, -2\pi)$, with
	e.g. $\psi = -2\pi$ corresponding to a grading by $(-1)^F e^{-2\pi i r}$.  
	
	Our large-coupling results are less complete than our results at small
	coupling, because we do not know how to exclude the possibility that there
	are some presently-unknown black objects which correspond to partially
	confined phases at large coupling. One might guess that
	increasing the 't Hooft coupling should increase eigenvalue fluctuations,
	and heuristically this could be expected to increase the deconfinement
	temperature and (naively) wipe out partially confined phases.  This 
	simple picture is consistent with our results. However,
	Ref.~\cite{ArabiArdehali:2019orz} recently also found evidence for
	partially-confined phases within the superconformal index, which is
	independent of the 't Hooft coupling.  So from the field theory point of
	view, partially confined phases appear to be fairly ubiquitous.  It would be
	very interesting to find new black objects dual to partially-confined
	phases, or to prove that they do not exist at least within the supergravity
	approximation.  

	{\bf Acknowledgements.}  We are grateful to M.~\"Unsal for conversations
	that inspired this work and collaboration at an initial stage, as well as to
	E.~Poppitz and J.~Sonner for helpful discussions.  A.~C. is supported by
	startup funds from the University of Minnesota, and A.~D. is supported by
	the National Science Foundation Graduate Research Fellowship under Grant No.
	00039202.

\bibliographystyle{apsrev4-1}
\bibliography{small_circle}

\begin{thebibliography}{65}%
\makeatletter
\providecommand \@ifxundefined [1]{%
 \@ifx{#1\undefined}
}%
\providecommand \@ifnum [1]{%
 \ifnum #1\expandafter \@firstoftwo
 \else \expandafter \@secondoftwo
 \fi
}%
\providecommand \@ifx [1]{%
 \ifx #1\expandafter \@firstoftwo
 \else \expandafter \@secondoftwo
 \fi
}%
\providecommand \natexlab [1]{#1}%
\providecommand \enquote  [1]{``#1''}%
\providecommand \bibnamefont  [1]{#1}%
\providecommand \bibfnamefont [1]{#1}%
\providecommand \citenamefont [1]{#1}%
\providecommand \href@noop [0]{\@secondoftwo}%
\providecommand \href [0]{\begingroup \@sanitize@url \@href}%
\providecommand \@href[1]{\@@startlink{#1}\@@href}%
\providecommand \@@href[1]{\endgroup#1\@@endlink}%
\providecommand \@sanitize@url [0]{\catcode `\\12\catcode `\$12\catcode
  `\&12\catcode `\#12\catcode `\^12\catcode `\_12\catcode `\%12\relax}%
\providecommand \@@startlink[1]{}%
\providecommand \@@endlink[0]{}%
\providecommand \url  [0]{\begingroup\@sanitize@url \@url }%
\providecommand \@url [1]{\endgroup\@href {#1}{\urlprefix }}%
\providecommand \urlprefix  [0]{URL }%
\providecommand \Eprint [0]{\href }%
\providecommand \doibase [0]{http://dx.doi.org/}%
\providecommand \selectlanguage [0]{\@gobble}%
\providecommand \bibinfo  [0]{\@secondoftwo}%
\providecommand \bibfield  [0]{\@secondoftwo}%
\providecommand \translation [1]{[#1]}%
\providecommand \BibitemOpen [0]{}%
\providecommand \bibitemStop [0]{}%
\providecommand \bibitemNoStop [0]{.\EOS\space}%
\providecommand \EOS [0]{\spacefactor3000\relax}%
\providecommand \BibitemShut  [1]{\csname bibitem#1\endcsname}%
\let\auto@bib@innerbib\@empty
\bibitem [{\citenamefont {Gross}\ \emph {et~al.}(1981)\citenamefont {Gross},
  \citenamefont {Pisarski},\ and\ \citenamefont {Yaffe}}]{Gross:1980br}%
  \BibitemOpen
  \bibfield  {author} {\bibinfo {author} {\bibfnamefont {D.~J.}\ \bibnamefont
  {Gross}}, \bibinfo {author} {\bibfnamefont {R.~D.}\ \bibnamefont {Pisarski}},
  \ and\ \bibinfo {author} {\bibfnamefont {L.~G.}\ \bibnamefont {Yaffe}},\
  }\href {\doibase 10.1103/RevModPhys.53.43} {\bibfield  {journal} {\bibinfo
  {journal} {Rev. Mod. Phys.}\ }\textbf {\bibinfo {volume} {53}},\ \bibinfo
  {pages} {43} (\bibinfo {year} {1981})}\BibitemShut {NoStop}%
\bibitem [{\citenamefont {Gaiotto}\ \emph {et~al.}(2015)\citenamefont
  {Gaiotto}, \citenamefont {Kapustin}, \citenamefont {Seiberg},\ and\
  \citenamefont {Willett}}]{Gaiotto:2014kfa}%
  \BibitemOpen
  \bibfield  {author} {\bibinfo {author} {\bibfnamefont {D.}~\bibnamefont
  {Gaiotto}}, \bibinfo {author} {\bibfnamefont {A.}~\bibnamefont {Kapustin}},
  \bibinfo {author} {\bibfnamefont {N.}~\bibnamefont {Seiberg}}, \ and\
  \bibinfo {author} {\bibfnamefont {B.}~\bibnamefont {Willett}},\ }\href
  {\doibase 10.1007/JHEP02(2015)172} {\bibfield  {journal} {\bibinfo  {journal}
  {JHEP}\ }\textbf {\bibinfo {volume} {02}},\ \bibinfo {pages} {172} (\bibinfo
  {year} {2015})},\ \Eprint {http://arxiv.org/abs/1412.5148} {arXiv:1412.5148
  [hep-th]} \BibitemShut {NoStop}%
\bibitem [{\citenamefont {Greensite}(2003)}]{Greensite:2003bk}%
  \BibitemOpen
  \bibfield  {author} {\bibinfo {author} {\bibfnamefont {J.}~\bibnamefont
  {Greensite}},\ }\href {\doibase 10.1016/S0146-6410(03)90012-3} {\bibfield
  {journal} {\bibinfo  {journal} {Prog. Part. Nucl. Phys.}\ }\textbf {\bibinfo
  {volume} {51}},\ \bibinfo {pages} {1} (\bibinfo {year} {2003})},\ \Eprint
  {http://arxiv.org/abs/hep-lat/0301023} {arXiv:hep-lat/0301023} \BibitemShut
  {NoStop}%
\bibitem [{\citenamefont {Sundborg}(2000)}]{Sundborg:1999ue}%
  \BibitemOpen
  \bibfield  {author} {\bibinfo {author} {\bibfnamefont {B.}~\bibnamefont
  {Sundborg}},\ }\href {\doibase 10.1016/S0550-3213(00)00044-4} {\bibfield
  {journal} {\bibinfo  {journal} {Nucl. Phys.}\ }\textbf {\bibinfo {volume}
  {B573}},\ \bibinfo {pages} {349} (\bibinfo {year} {2000})},\ \Eprint
  {http://arxiv.org/abs/hep-th/9908001} {arXiv:hep-th/9908001 [hep-th]}
  \BibitemShut {NoStop}%
\bibitem [{\citenamefont {Aharony}\ \emph {et~al.}(2004)\citenamefont
  {Aharony}, \citenamefont {Marsano}, \citenamefont {Minwalla}, \citenamefont
  {Papadodimas},\ and\ \citenamefont {Van~Raamsdonk}}]{Aharony:2003sx}%
  \BibitemOpen
  \bibfield  {author} {\bibinfo {author} {\bibfnamefont {O.}~\bibnamefont
  {Aharony}}, \bibinfo {author} {\bibfnamefont {J.}~\bibnamefont {Marsano}},
  \bibinfo {author} {\bibfnamefont {S.}~\bibnamefont {Minwalla}}, \bibinfo
  {author} {\bibfnamefont {K.}~\bibnamefont {Papadodimas}}, \ and\ \bibinfo
  {author} {\bibfnamefont {M.}~\bibnamefont {Van~Raamsdonk}},\ }\bibfield
  {booktitle} {\emph {\bibinfo {booktitle} {{Lie theory and its applications in
  physics. Proceedings, 5th International Workshop, Varna, Bulgaria, June
  16-22, 2003}}},\ }\href {\doibase 10.4310/ATMP.2004.v8.n4.a1} {\bibfield
  {journal} {\bibinfo  {journal} {Adv. Theor. Math. Phys.}\ }\textbf {\bibinfo
  {volume} {8}},\ \bibinfo {pages} {603} (\bibinfo {year} {2004})},\ \Eprint
  {http://arxiv.org/abs/hep-th/0310285} {arXiv:hep-th/0310285 [hep-th]}
  \BibitemShut {NoStop}%
\bibitem [{\citenamefont {Aharony}\ \emph {et~al.}(2005)\citenamefont
  {Aharony}, \citenamefont {Marsano}, \citenamefont {Minwalla}, \citenamefont
  {Papadodimas},\ and\ \citenamefont {Van~Raamsdonk}}]{Aharony:2005bq}%
  \BibitemOpen
  \bibfield  {author} {\bibinfo {author} {\bibfnamefont {O.}~\bibnamefont
  {Aharony}}, \bibinfo {author} {\bibfnamefont {J.}~\bibnamefont {Marsano}},
  \bibinfo {author} {\bibfnamefont {S.}~\bibnamefont {Minwalla}}, \bibinfo
  {author} {\bibfnamefont {K.}~\bibnamefont {Papadodimas}}, \ and\ \bibinfo
  {author} {\bibfnamefont {M.}~\bibnamefont {Van~Raamsdonk}},\ }\href {\doibase
  10.1103/PhysRevD.71.125018} {\bibfield  {journal} {\bibinfo  {journal} {Phys.
  Rev.}\ }\textbf {\bibinfo {volume} {D71}},\ \bibinfo {pages} {125018}
  (\bibinfo {year} {2005})},\ \Eprint {http://arxiv.org/abs/hep-th/0502149}
  {arXiv:hep-th/0502149 [hep-th]} \BibitemShut {NoStop}%
\bibitem [{\citenamefont {Kovtun}\ \emph {et~al.}(2007)\citenamefont {Kovtun},
  \citenamefont {Unsal},\ and\ \citenamefont {Yaffe}}]{Kovtun:2007py}%
  \BibitemOpen
  \bibfield  {author} {\bibinfo {author} {\bibfnamefont {P.}~\bibnamefont
  {Kovtun}}, \bibinfo {author} {\bibfnamefont {M.}~\bibnamefont {Unsal}}, \
  and\ \bibinfo {author} {\bibfnamefont {L.~G.}\ \bibnamefont {Yaffe}},\ }\href
  {\doibase 10.1088/1126-6708/2007/06/019} {\bibfield  {journal} {\bibinfo
  {journal} {JHEP}\ }\textbf {\bibinfo {volume} {06}},\ \bibinfo {pages} {019}
  (\bibinfo {year} {2007})},\ \Eprint {http://arxiv.org/abs/hep-th/0702021}
  {arXiv:hep-th/0702021 [HEP-TH]} \BibitemShut {NoStop}%
\bibitem [{\citenamefont {Basar}\ \emph {et~al.}(2013)\citenamefont {Basar},
  \citenamefont {Cherman}, \citenamefont {Dorigoni},\ and\ \citenamefont
  {Unsal}}]{Basar:2013sza}%
  \BibitemOpen
  \bibfield  {author} {\bibinfo {author} {\bibfnamefont {G.}~\bibnamefont
  {Basar}}, \bibinfo {author} {\bibfnamefont {A.}~\bibnamefont {Cherman}},
  \bibinfo {author} {\bibfnamefont {D.}~\bibnamefont {Dorigoni}}, \ and\
  \bibinfo {author} {\bibfnamefont {M.}~\bibnamefont {Unsal}},\ }\href
  {\doibase 10.1103/PhysRevLett.111.121601} {\bibfield  {journal} {\bibinfo
  {journal} {Phys. Rev. Lett. 111,}\ }\textbf {\bibinfo {volume} {121601}},\
  \bibinfo {pages} {121601} (\bibinfo {year} {2013})},\ \Eprint
  {http://arxiv.org/abs/1306.2960} {arXiv:1306.2960 [hep-th]} \BibitemShut
  {NoStop}%
\bibitem [{\citenamefont {Basar}\ \emph {et~al.}(2015)\citenamefont {Basar},
  \citenamefont {Cherman},\ and\ \citenamefont {McGady}}]{Basar:2014jua}%
  \BibitemOpen
  \bibfield  {author} {\bibinfo {author} {\bibfnamefont {G.}~\bibnamefont
  {Basar}}, \bibinfo {author} {\bibfnamefont {A.}~\bibnamefont {Cherman}}, \
  and\ \bibinfo {author} {\bibfnamefont {D.~A.}\ \bibnamefont {McGady}},\
  }\href {\doibase http://dx.doi.org/10.1007/JHEP07(2015)016,
  10.1007/JHEP07(2015)016} {\bibfield  {journal} {\bibinfo  {journal} {JHEP}\
  }\textbf {\bibinfo {volume} {07}},\ \bibinfo {pages} {016} (\bibinfo {year}
  {2015})},\ \Eprint {http://arxiv.org/abs/1409.1617} {arXiv:1409.1617
  [hep-th]} \BibitemShut {NoStop}%
\bibitem [{\citenamefont {Cherman}\ \emph {et~al.}(2019)\citenamefont
  {Cherman}, \citenamefont {Shifman},\ and\ \citenamefont
  {{\"U}nsal}}]{Cherman:2018mya}%
  \BibitemOpen
  \bibfield  {author} {\bibinfo {author} {\bibfnamefont {A.}~\bibnamefont
  {Cherman}}, \bibinfo {author} {\bibfnamefont {M.}~\bibnamefont {Shifman}}, \
  and\ \bibinfo {author} {\bibfnamefont {M.}~\bibnamefont {{\"U}nsal}},\ }\href
  {\doibase 10.1103/PhysRevD.99.105001} {\bibfield  {journal} {\bibinfo
  {journal} {Phys. Rev.}\ }\textbf {\bibinfo {volume} {D99}},\ \bibinfo {pages}
  {105001} (\bibinfo {year} {2019})},\ \Eprint
  {http://arxiv.org/abs/1812.04642} {arXiv:1812.04642 [hep-th]} \BibitemShut
  {NoStop}%
\bibitem [{\citenamefont {Cherman}\ \emph {et~al.}(2020)\citenamefont
  {Cherman}, \citenamefont {Kamata}, \citenamefont {Sch\"afer},\ and\
  \citenamefont {\"Unsal}}]{Cherman:2019ecx}%
  \BibitemOpen
  \bibfield  {author} {\bibinfo {author} {\bibfnamefont {A.}~\bibnamefont
  {Cherman}}, \bibinfo {author} {\bibfnamefont {S.}~\bibnamefont {Kamata}},
  \bibinfo {author} {\bibfnamefont {T.}~\bibnamefont {Sch\"afer}}, \ and\
  \bibinfo {author} {\bibfnamefont {M.}~\bibnamefont {\"Unsal}},\ }\href
  {\doibase 10.1103/PhysRevD.101.014012} {\bibfield  {journal} {\bibinfo
  {journal} {Phys. Rev. D}\ }\textbf {\bibinfo {volume} {101}},\ \bibinfo
  {pages} {014012} (\bibinfo {year} {2020})},\ \Eprint
  {http://arxiv.org/abs/1910.12312} {arXiv:1910.12312 [hep-th]} \BibitemShut
  {NoStop}%
\bibitem [{\citenamefont {Kanazawa}\ and\ \citenamefont
  {\"Unsal}(2020)}]{Kanazawa:2019tnf}%
  \BibitemOpen
  \bibfield  {author} {\bibinfo {author} {\bibfnamefont {T.}~\bibnamefont
  {Kanazawa}}\ and\ \bibinfo {author} {\bibfnamefont {M.}~\bibnamefont
  {\"Unsal}},\ }\href {\doibase 10.1103/PhysRevD.102.034013} {\bibfield
  {journal} {\bibinfo  {journal} {Phys. Rev. D}\ }\textbf {\bibinfo {volume}
  {102}},\ \bibinfo {pages} {034013} (\bibinfo {year} {2020})},\ \Eprint
  {http://arxiv.org/abs/1909.05222} {arXiv:1909.05222 [hep-th]} \BibitemShut
  {NoStop}%
\bibitem [{\citenamefont {Maldacena}(1999)}]{Maldacena:1997re}%
  \BibitemOpen
  \bibfield  {author} {\bibinfo {author} {\bibfnamefont {J.~M.}\ \bibnamefont
  {Maldacena}},\ }\href {\doibase 10.1023/A:1026654312961,
  10.4310/ATMP.1998.v2.n2.a1} {\bibfield  {journal} {\bibinfo  {journal} {Int.
  J. Theor. Phys.}\ }\textbf {\bibinfo {volume} {38}},\ \bibinfo {pages} {1113}
  (\bibinfo {year} {1999})},\ \bibinfo {note} {[Adv. Theor. Math.
  Phys.2,231(1998)]},\ \Eprint {http://arxiv.org/abs/hep-th/9711200}
  {arXiv:hep-th/9711200 [hep-th]} \BibitemShut {NoStop}%
\bibitem [{\citenamefont {Gubser}\ \emph {et~al.}(1998)\citenamefont {Gubser},
  \citenamefont {Klebanov},\ and\ \citenamefont {Polyakov}}]{Gubser:1998bc}%
  \BibitemOpen
  \bibfield  {author} {\bibinfo {author} {\bibfnamefont {S.~S.}\ \bibnamefont
  {Gubser}}, \bibinfo {author} {\bibfnamefont {I.~R.}\ \bibnamefont
  {Klebanov}}, \ and\ \bibinfo {author} {\bibfnamefont {A.~M.}\ \bibnamefont
  {Polyakov}},\ }\href {\doibase 10.1016/S0370-2693(98)00377-3} {\bibfield
  {journal} {\bibinfo  {journal} {Phys. Lett.}\ }\textbf {\bibinfo {volume}
  {B428}},\ \bibinfo {pages} {105} (\bibinfo {year} {1998})},\ \Eprint
  {http://arxiv.org/abs/hep-th/9802109} {arXiv:hep-th/9802109 [hep-th]}
  \BibitemShut {NoStop}%
\bibitem [{\citenamefont {Witten}(1998{\natexlab{a}})}]{Witten:1998qj}%
  \BibitemOpen
  \bibfield  {author} {\bibinfo {author} {\bibfnamefont {E.}~\bibnamefont
  {Witten}},\ }\href {\doibase 10.4310/ATMP.1998.v2.n2.a2} {\bibfield
  {journal} {\bibinfo  {journal} {Adv. Theor. Math. Phys.}\ }\textbf {\bibinfo
  {volume} {2}},\ \bibinfo {pages} {253} (\bibinfo {year}
  {1998}{\natexlab{a}})},\ \Eprint {http://arxiv.org/abs/hep-th/9802150}
  {arXiv:hep-th/9802150 [hep-th]} \BibitemShut {NoStop}%
\bibitem [{\citenamefont {Witten}(1998{\natexlab{b}})}]{Witten:1998zw}%
  \BibitemOpen
  \bibfield  {author} {\bibinfo {author} {\bibfnamefont {E.}~\bibnamefont
  {Witten}},\ }\href@noop {} {\bibfield  {journal} {\bibinfo  {journal} {Adv.
  Theor. Math. Phys.}\ }\textbf {\bibinfo {volume} {2}},\ \bibinfo {pages}
  {505} (\bibinfo {year} {1998}{\natexlab{b}})},\ \Eprint
  {http://arxiv.org/abs/hep-th/9803131} {arXiv:hep-th/9803131 [hep-th]}
  \BibitemShut {NoStop}%
\bibitem [{\citenamefont {Harmark}\ and\ \citenamefont
  {Wilhelm}(2018{\natexlab{a}})}]{Harmark:2017yrv}%
  \BibitemOpen
  \bibfield  {author} {\bibinfo {author} {\bibfnamefont {T.}~\bibnamefont
  {Harmark}}\ and\ \bibinfo {author} {\bibfnamefont {M.}~\bibnamefont
  {Wilhelm}},\ }\href {\doibase 10.1103/PhysRevLett.120.071605} {\bibfield
  {journal} {\bibinfo  {journal} {Phys. Rev. Lett.}\ }\textbf {\bibinfo
  {volume} {120}},\ \bibinfo {pages} {071605} (\bibinfo {year}
  {2018}{\natexlab{a}})},\ \Eprint {http://arxiv.org/abs/1706.03074}
  {arXiv:1706.03074 [hep-th]} \BibitemShut {NoStop}%
\bibitem [{\citenamefont {Harmark}\ and\ \citenamefont
  {Wilhelm}(2018{\natexlab{b}})}]{Harmark:2018red}%
  \BibitemOpen
  \bibfield  {author} {\bibinfo {author} {\bibfnamefont {T.}~\bibnamefont
  {Harmark}}\ and\ \bibinfo {author} {\bibfnamefont {M.}~\bibnamefont
  {Wilhelm}},\ }\href {\doibase 10.1016/j.physletb.2018.09.033} {\bibfield
  {journal} {\bibinfo  {journal} {Phys. Lett.}\ }\textbf {\bibinfo {volume}
  {B786}},\ \bibinfo {pages} {53} (\bibinfo {year} {2018}{\natexlab{b}})},\
  \Eprint {http://arxiv.org/abs/1803.04416} {arXiv:1803.04416 [hep-th]}
  \BibitemShut {NoStop}%
\bibitem [{\citenamefont {Yamada}\ and\ \citenamefont
  {Yaffe}(2006)}]{Yamada:2006rx}%
  \BibitemOpen
  \bibfield  {author} {\bibinfo {author} {\bibfnamefont {D.}~\bibnamefont
  {Yamada}}\ and\ \bibinfo {author} {\bibfnamefont {L.~G.}\ \bibnamefont
  {Yaffe}},\ }\href {\doibase 10.1088/1126-6708/2006/09/027} {\bibfield
  {journal} {\bibinfo  {journal} {JHEP}\ }\textbf {\bibinfo {volume} {09}},\
  \bibinfo {pages} {027} (\bibinfo {year} {2006})},\ \Eprint
  {http://arxiv.org/abs/hep-th/0602074} {arXiv:hep-th/0602074 [hep-th]}
  \BibitemShut {NoStop}%
\bibitem [{\citenamefont {Benini}\ and\ \citenamefont
  {Milan}(2020{\natexlab{a}})}]{Benini:2018ywd}%
  \BibitemOpen
  \bibfield  {author} {\bibinfo {author} {\bibfnamefont {F.}~\bibnamefont
  {Benini}}\ and\ \bibinfo {author} {\bibfnamefont {P.}~\bibnamefont {Milan}},\
  }\href {\doibase 10.1103/PhysRevX.10.021037} {\bibfield  {journal} {\bibinfo
  {journal} {Phys. Rev. X}\ }\textbf {\bibinfo {volume} {10}},\ \bibinfo
  {pages} {021037} (\bibinfo {year} {2020}{\natexlab{a}})},\ \Eprint
  {http://arxiv.org/abs/1812.09613} {arXiv:1812.09613 [hep-th]} \BibitemShut
  {NoStop}%
\bibitem [{\citenamefont {Copetti}\ \emph {et~al.}(2020)\citenamefont
  {Copetti}, \citenamefont {Grassi}, \citenamefont {Komargodski},\ and\
  \citenamefont {Tizzano}}]{Copetti:2020dil}%
  \BibitemOpen
  \bibfield  {author} {\bibinfo {author} {\bibfnamefont {C.}~\bibnamefont
  {Copetti}}, \bibinfo {author} {\bibfnamefont {A.}~\bibnamefont {Grassi}},
  \bibinfo {author} {\bibfnamefont {Z.}~\bibnamefont {Komargodski}}, \ and\
  \bibinfo {author} {\bibfnamefont {L.}~\bibnamefont {Tizzano}},\ }\href@noop
  {} {\  (\bibinfo {year} {2020})},\ \Eprint {http://arxiv.org/abs/2008.04950}
  {arXiv:2008.04950 [hep-th]} \BibitemShut {NoStop}%
\bibitem [{\citenamefont {Myers}\ and\ \citenamefont
  {Ogilvie}(2008)}]{Myers:2007vc}%
  \BibitemOpen
  \bibfield  {author} {\bibinfo {author} {\bibfnamefont {J.~C.}\ \bibnamefont
  {Myers}}\ and\ \bibinfo {author} {\bibfnamefont {M.~C.}\ \bibnamefont
  {Ogilvie}},\ }\href {\doibase 10.1103/PhysRevD.77.125030} {\bibfield
  {journal} {\bibinfo  {journal} {Phys. Rev.}\ }\textbf {\bibinfo {volume}
  {D77}},\ \bibinfo {pages} {125030} (\bibinfo {year} {2008})},\ \Eprint
  {http://arxiv.org/abs/0707.1869} {arXiv:0707.1869 [hep-lat]} \BibitemShut
  {NoStop}%
\bibitem [{\citenamefont {Mussel}\ and\ \citenamefont
  {Yacoby}(2009)}]{Mussel:2009uw}%
  \BibitemOpen
  \bibfield  {author} {\bibinfo {author} {\bibfnamefont {M.}~\bibnamefont
  {Mussel}}\ and\ \bibinfo {author} {\bibfnamefont {R.}~\bibnamefont
  {Yacoby}},\ }\href {\doibase 10.1088/1126-6708/2009/12/005} {\bibfield
  {journal} {\bibinfo  {journal} {JHEP}\ }\textbf {\bibinfo {volume} {12}},\
  \bibinfo {pages} {005} (\bibinfo {year} {2009})},\ \Eprint
  {http://arxiv.org/abs/0909.0407} {arXiv:0909.0407 [hep-th]} \BibitemShut
  {NoStop}%
\bibitem [{\citenamefont {Chamblin}\ \emph
  {et~al.}(1999{\natexlab{a}})\citenamefont {Chamblin}, \citenamefont
  {Emparan}, \citenamefont {Johnson},\ and\ \citenamefont
  {Myers}}]{Chamblin:1999tk}%
  \BibitemOpen
  \bibfield  {author} {\bibinfo {author} {\bibfnamefont {A.}~\bibnamefont
  {Chamblin}}, \bibinfo {author} {\bibfnamefont {R.}~\bibnamefont {Emparan}},
  \bibinfo {author} {\bibfnamefont {C.~V.}\ \bibnamefont {Johnson}}, \ and\
  \bibinfo {author} {\bibfnamefont {R.~C.}\ \bibnamefont {Myers}},\ }\href
  {\doibase 10.1103/PhysRevD.60.064018} {\bibfield  {journal} {\bibinfo
  {journal} {Phys.Rev.D}\ }\textbf {\bibinfo {volume} {60}},\ \bibinfo {pages}
  {064018} (\bibinfo {year} {1999}{\natexlab{a}})},\ \Eprint
  {http://arxiv.org/abs/hep-th/9902170} {arXiv:hep-th/9902170} \BibitemShut
  {NoStop}%
\bibitem [{\citenamefont {Chamblin}\ \emph
  {et~al.}(1999{\natexlab{b}})\citenamefont {Chamblin}, \citenamefont
  {Emparan}, \citenamefont {Johnson},\ and\ \citenamefont
  {Myers}}]{Chamblin:1999hg}%
  \BibitemOpen
  \bibfield  {author} {\bibinfo {author} {\bibfnamefont {A.}~\bibnamefont
  {Chamblin}}, \bibinfo {author} {\bibfnamefont {R.}~\bibnamefont {Emparan}},
  \bibinfo {author} {\bibfnamefont {C.~V.}\ \bibnamefont {Johnson}}, \ and\
  \bibinfo {author} {\bibfnamefont {R.~C.}\ \bibnamefont {Myers}},\ }\href
  {\doibase 10.1103/PhysRevD.60.104026} {\bibfield  {journal} {\bibinfo
  {journal} {Phys. Rev. D}\ }\textbf {\bibinfo {volume} {60}},\ \bibinfo
  {pages} {104026} (\bibinfo {year} {1999}{\natexlab{b}})},\ \Eprint
  {http://arxiv.org/abs/hep-th/9904197} {arXiv:hep-th/9904197} \BibitemShut
  {NoStop}%
\bibitem [{\citenamefont {Chong}\ \emph
  {et~al.}(2005{\natexlab{a}})\citenamefont {Chong}, \citenamefont {Cvetic},
  \citenamefont {Lu},\ and\ \citenamefont {Pope}}]{Chong:2005da}%
  \BibitemOpen
  \bibfield  {author} {\bibinfo {author} {\bibfnamefont {Z.}~\bibnamefont
  {Chong}}, \bibinfo {author} {\bibfnamefont {M.}~\bibnamefont {Cvetic}},
  \bibinfo {author} {\bibfnamefont {H.}~\bibnamefont {Lu}}, \ and\ \bibinfo
  {author} {\bibfnamefont {C.}~\bibnamefont {Pope}},\ }\href {\doibase
  10.1103/PhysRevD.72.041901} {\bibfield  {journal} {\bibinfo  {journal} {Phys.
  Rev. D}\ }\textbf {\bibinfo {volume} {72}},\ \bibinfo {pages} {041901}
  (\bibinfo {year} {2005}{\natexlab{a}})},\ \Eprint
  {http://arxiv.org/abs/hep-th/0505112} {arXiv:hep-th/0505112} \BibitemShut
  {NoStop}%
\bibitem [{\citenamefont {Chong}\ \emph {et~al.}(2007)\citenamefont {Chong},
  \citenamefont {Cvetic}, \citenamefont {Lu},\ and\ \citenamefont
  {Pope}}]{Chong:2006zx}%
  \BibitemOpen
  \bibfield  {author} {\bibinfo {author} {\bibfnamefont {Z.}~\bibnamefont
  {Chong}}, \bibinfo {author} {\bibfnamefont {M.}~\bibnamefont {Cvetic}},
  \bibinfo {author} {\bibfnamefont {H.}~\bibnamefont {Lu}}, \ and\ \bibinfo
  {author} {\bibfnamefont {C.}~\bibnamefont {Pope}},\ }\href {\doibase
  10.1016/j.physletb.2006.11.012} {\bibfield  {journal} {\bibinfo  {journal}
  {Phys. Lett. B}\ }\textbf {\bibinfo {volume} {644}},\ \bibinfo {pages} {192}
  (\bibinfo {year} {2007})},\ \Eprint {http://arxiv.org/abs/hep-th/0606213}
  {arXiv:hep-th/0606213} \BibitemShut {NoStop}%
\bibitem [{\citenamefont {Chong}\ \emph
  {et~al.}(2005{\natexlab{b}})\citenamefont {Chong}, \citenamefont {Cvetic},
  \citenamefont {Lu},\ and\ \citenamefont {Pope}}]{Chong:2005hr}%
  \BibitemOpen
  \bibfield  {author} {\bibinfo {author} {\bibfnamefont {Z.-W.}\ \bibnamefont
  {Chong}}, \bibinfo {author} {\bibfnamefont {M.}~\bibnamefont {Cvetic}},
  \bibinfo {author} {\bibfnamefont {H.}~\bibnamefont {Lu}}, \ and\ \bibinfo
  {author} {\bibfnamefont {C.}~\bibnamefont {Pope}},\ }\href {\doibase
  10.1103/PhysRevLett.95.161301} {\bibfield  {journal} {\bibinfo  {journal}
  {Phys. Rev. Lett.}\ }\textbf {\bibinfo {volume} {95}},\ \bibinfo {pages}
  {161301} (\bibinfo {year} {2005}{\natexlab{b}})},\ \Eprint
  {http://arxiv.org/abs/hep-th/0506029} {arXiv:hep-th/0506029} \BibitemShut
  {NoStop}%
\bibitem [{\citenamefont {Cvetic}\ and\ \citenamefont
  {Gubser}(1999)}]{Cvetic:1999ne}%
  \BibitemOpen
  \bibfield  {author} {\bibinfo {author} {\bibfnamefont {M.}~\bibnamefont
  {Cvetic}}\ and\ \bibinfo {author} {\bibfnamefont {S.~S.}\ \bibnamefont
  {Gubser}},\ }\href {\doibase 10.1088/1126-6708/1999/04/024} {\bibfield
  {journal} {\bibinfo  {journal} {JHEP}\ }\textbf {\bibinfo {volume} {04}},\
  \bibinfo {pages} {024} (\bibinfo {year} {1999})},\ \Eprint
  {http://arxiv.org/abs/hep-th/9902195} {arXiv:hep-th/9902195 [hep-th]}
  \BibitemShut {NoStop}%
\bibitem [{\citenamefont {Cvetic}\ \emph {et~al.}(2004)\citenamefont {Cvetic},
  \citenamefont {Lu},\ and\ \citenamefont {Pope}}]{Cvetic:2004ny}%
  \BibitemOpen
  \bibfield  {author} {\bibinfo {author} {\bibfnamefont {M.}~\bibnamefont
  {Cvetic}}, \bibinfo {author} {\bibfnamefont {H.}~\bibnamefont {Lu}}, \ and\
  \bibinfo {author} {\bibfnamefont {C.}~\bibnamefont {Pope}},\ }\href {\doibase
  10.1103/PhysRevD.70.081502} {\bibfield  {journal} {\bibinfo  {journal} {Phys.
  Rev. D}\ }\textbf {\bibinfo {volume} {70}},\ \bibinfo {pages} {081502}
  (\bibinfo {year} {2004})},\ \Eprint {http://arxiv.org/abs/hep-th/0407058}
  {arXiv:hep-th/0407058} \BibitemShut {NoStop}%
\bibitem [{\citenamefont {Emparan}\ \emph {et~al.}(1999)\citenamefont
  {Emparan}, \citenamefont {Johnson},\ and\ \citenamefont
  {Myers}}]{Emparan:1999pm}%
  \BibitemOpen
  \bibfield  {author} {\bibinfo {author} {\bibfnamefont {R.}~\bibnamefont
  {Emparan}}, \bibinfo {author} {\bibfnamefont {C.~V.}\ \bibnamefont
  {Johnson}}, \ and\ \bibinfo {author} {\bibfnamefont {R.~C.}\ \bibnamefont
  {Myers}},\ }\href {\doibase 10.1103/PhysRevD.60.104001} {\bibfield  {journal}
  {\bibinfo  {journal} {Phys. Rev.}\ }\textbf {\bibinfo {volume} {D60}},\
  \bibinfo {pages} {104001} (\bibinfo {year} {1999})},\ \Eprint
  {http://arxiv.org/abs/hep-th/9903238} {arXiv:hep-th/9903238 [hep-th]}
  \BibitemShut {NoStop}%
\bibitem [{\citenamefont {Hawking}\ and\ \citenamefont
  {Page}(1983)}]{Hawking:1982dh}%
  \BibitemOpen
  \bibfield  {author} {\bibinfo {author} {\bibfnamefont {S.~W.}\ \bibnamefont
  {Hawking}}\ and\ \bibinfo {author} {\bibfnamefont {D.~N.}\ \bibnamefont
  {Page}},\ }\href {\doibase 10.1007/BF01208266} {\bibfield  {journal}
  {\bibinfo  {journal} {Commun. Math. Phys.}\ }\textbf {\bibinfo {volume}
  {87}},\ \bibinfo {pages} {577} (\bibinfo {year} {1983})}\BibitemShut
  {NoStop}%
\bibitem [{\citenamefont {Aharony}\ and\ \citenamefont
  {Witten}(1998)}]{Aharony:1998qu}%
  \BibitemOpen
  \bibfield  {author} {\bibinfo {author} {\bibfnamefont {O.}~\bibnamefont
  {Aharony}}\ and\ \bibinfo {author} {\bibfnamefont {E.}~\bibnamefont
  {Witten}},\ }\href {\doibase 10.1088/1126-6708/1998/11/018} {\bibfield
  {journal} {\bibinfo  {journal} {JHEP}\ }\textbf {\bibinfo {volume} {11}},\
  \bibinfo {pages} {018} (\bibinfo {year} {1998})},\ \Eprint
  {http://arxiv.org/abs/hep-th/9807205} {arXiv:hep-th/9807205} \BibitemShut
  {NoStop}%
\bibitem [{\citenamefont {Aarts}\ \emph {et~al.}(2010)\citenamefont {Aarts},
  \citenamefont {Kumar},\ and\ \citenamefont {Rafferty}}]{Aarts:2010ky}%
  \BibitemOpen
  \bibfield  {author} {\bibinfo {author} {\bibfnamefont {G.}~\bibnamefont
  {Aarts}}, \bibinfo {author} {\bibfnamefont {S.}~\bibnamefont {Kumar}}, \ and\
  \bibinfo {author} {\bibfnamefont {J.}~\bibnamefont {Rafferty}},\ }\href
  {\doibase 10.1007/JHEP07(2010)056} {\bibfield  {journal} {\bibinfo  {journal}
  {JHEP}\ }\textbf {\bibinfo {volume} {07}},\ \bibinfo {pages} {056} (\bibinfo
  {year} {2010})},\ \Eprint {http://arxiv.org/abs/1005.2947} {arXiv:1005.2947
  [hep-th]} \BibitemShut {NoStop}%
\bibitem [{\citenamefont {Hanada}\ \emph
  {et~al.}(2019{\natexlab{a}})\citenamefont {Hanada}, \citenamefont {Ishiki},\
  and\ \citenamefont {Watanabe}}]{Hanada:2018zxn}%
  \BibitemOpen
  \bibfield  {author} {\bibinfo {author} {\bibfnamefont {M.}~\bibnamefont
  {Hanada}}, \bibinfo {author} {\bibfnamefont {G.}~\bibnamefont {Ishiki}}, \
  and\ \bibinfo {author} {\bibfnamefont {H.}~\bibnamefont {Watanabe}},\ }\href
  {\doibase 10.1007/JHEP03(2019)145} {\bibfield  {journal} {\bibinfo  {journal}
  {JHEP}\ }\textbf {\bibinfo {volume} {03}},\ \bibinfo {pages} {145} (\bibinfo
  {year} {2019}{\natexlab{a}})},\ \bibinfo {note} {[Erratum: JHEP 10, 029
  (2019)]},\ \Eprint {http://arxiv.org/abs/1812.05494} {arXiv:1812.05494
  [hep-th]} \BibitemShut {NoStop}%
\bibitem [{\citenamefont {Berenstein}(2018)}]{Berenstein:2018lrm}%
  \BibitemOpen
  \bibfield  {author} {\bibinfo {author} {\bibfnamefont {D.}~\bibnamefont
  {Berenstein}},\ }\href {\doibase 10.1007/JHEP09(2018)054} {\bibfield
  {journal} {\bibinfo  {journal} {JHEP}\ }\textbf {\bibinfo {volume} {09}},\
  \bibinfo {pages} {054} (\bibinfo {year} {2018})},\ \Eprint
  {http://arxiv.org/abs/1806.05729} {arXiv:1806.05729 [hep-th]} \BibitemShut
  {NoStop}%
\bibitem [{\citenamefont {Hanada}\ \emph
  {et~al.}(2019{\natexlab{b}})\citenamefont {Hanada}, \citenamefont {Jevicki},
  \citenamefont {Peng},\ and\ \citenamefont {Wintergerst}}]{Hanada:2019czd}%
  \BibitemOpen
  \bibfield  {author} {\bibinfo {author} {\bibfnamefont {M.}~\bibnamefont
  {Hanada}}, \bibinfo {author} {\bibfnamefont {A.}~\bibnamefont {Jevicki}},
  \bibinfo {author} {\bibfnamefont {C.}~\bibnamefont {Peng}}, \ and\ \bibinfo
  {author} {\bibfnamefont {N.}~\bibnamefont {Wintergerst}},\ }\href {\doibase
  10.1007/JHEP12(2019)167} {\bibfield  {journal} {\bibinfo  {journal} {JHEP}\
  }\textbf {\bibinfo {volume} {12}},\ \bibinfo {pages} {167} (\bibinfo {year}
  {2019}{\natexlab{b}})},\ \Eprint {http://arxiv.org/abs/1909.09118}
  {arXiv:1909.09118 [hep-th]} \BibitemShut {NoStop}%
\bibitem [{\citenamefont {Hanada}\ and\ \citenamefont
  {Robinson}(2020)}]{Hanada:2019kue}%
  \BibitemOpen
  \bibfield  {author} {\bibinfo {author} {\bibfnamefont {M.}~\bibnamefont
  {Hanada}}\ and\ \bibinfo {author} {\bibfnamefont {B.}~\bibnamefont
  {Robinson}},\ }\href {\doibase 10.1103/PhysRevD.102.096013} {\bibfield
  {journal} {\bibinfo  {journal} {Phys. Rev. D}\ }\textbf {\bibinfo {volume}
  {102}},\ \bibinfo {pages} {096013} (\bibinfo {year} {2020})},\ \Eprint
  {http://arxiv.org/abs/1911.06223} {arXiv:1911.06223 [hep-th]} \BibitemShut
  {NoStop}%
\bibitem [{\citenamefont {Hanada}\ \emph
  {et~al.}(2019{\natexlab{c}})\citenamefont {Hanada}, \citenamefont {Ishiki},\
  and\ \citenamefont {Watanabe}}]{Hanada:2019rzv}%
  \BibitemOpen
  \bibfield  {author} {\bibinfo {author} {\bibfnamefont {M.}~\bibnamefont
  {Hanada}}, \bibinfo {author} {\bibfnamefont {G.}~\bibnamefont {Ishiki}}, \
  and\ \bibinfo {author} {\bibfnamefont {H.}~\bibnamefont {Watanabe}},\ }\href
  {\doibase 10.22323/1.363.0055} {\bibfield  {journal} {\bibinfo  {journal}
  {PoS}\ }\textbf {\bibinfo {volume} {LATTICE2019}},\ \bibinfo {pages} {055}
  (\bibinfo {year} {2019}{\natexlab{c}})},\ \Eprint
  {http://arxiv.org/abs/1911.11465} {arXiv:1911.11465 [hep-lat]} \BibitemShut
  {NoStop}%
\bibitem [{\citenamefont {Hanada}\ \emph {et~al.}(2020)\citenamefont {Hanada},
  \citenamefont {Shimada},\ and\ \citenamefont {Wintergerst}}]{Hanada:2020uvt}%
  \BibitemOpen
  \bibfield  {author} {\bibinfo {author} {\bibfnamefont {M.}~\bibnamefont
  {Hanada}}, \bibinfo {author} {\bibfnamefont {H.}~\bibnamefont {Shimada}}, \
  and\ \bibinfo {author} {\bibfnamefont {N.}~\bibnamefont {Wintergerst}},\
  }\href@noop {} {\  (\bibinfo {year} {2020})},\ \Eprint
  {http://arxiv.org/abs/2001.10459} {arXiv:2001.10459 [hep-th]} \BibitemShut
  {NoStop}%
\bibitem [{\citenamefont {Watanabe}\ \emph {et~al.}(2020)\citenamefont
  {Watanabe}, \citenamefont {Bergner}, \citenamefont {Bodendorfer},
  \citenamefont {Shiba~Funai}, \citenamefont {Hanada}, \citenamefont {Rinaldi},
  \citenamefont {Sch\"afer},\ and\ \citenamefont {Vranas}}]{Watanabe:2020ufk}%
  \BibitemOpen
  \bibfield  {author} {\bibinfo {author} {\bibfnamefont {H.}~\bibnamefont
  {Watanabe}}, \bibinfo {author} {\bibfnamefont {G.}~\bibnamefont {Bergner}},
  \bibinfo {author} {\bibfnamefont {N.}~\bibnamefont {Bodendorfer}}, \bibinfo
  {author} {\bibfnamefont {S.}~\bibnamefont {Shiba~Funai}}, \bibinfo {author}
  {\bibfnamefont {M.}~\bibnamefont {Hanada}}, \bibinfo {author} {\bibfnamefont
  {E.}~\bibnamefont {Rinaldi}}, \bibinfo {author} {\bibfnamefont
  {A.}~\bibnamefont {Sch\"afer}}, \ and\ \bibinfo {author} {\bibfnamefont
  {P.}~\bibnamefont {Vranas}},\ }\href@noop {} {\  (\bibinfo {year} {2020})},\
  \Eprint {http://arxiv.org/abs/2005.04103} {arXiv:2005.04103 [hep-th]}
  \BibitemShut {NoStop}%
\bibitem [{\citenamefont {Arabi~Ardehali}\ \emph {et~al.}(2020)\citenamefont
  {Arabi~Ardehali}, \citenamefont {Hong},\ and\ \citenamefont
  {Liu}}]{ArabiArdehali:2019orz}%
  \BibitemOpen
  \bibfield  {author} {\bibinfo {author} {\bibfnamefont {A.}~\bibnamefont
  {Arabi~Ardehali}}, \bibinfo {author} {\bibfnamefont {J.}~\bibnamefont
  {Hong}}, \ and\ \bibinfo {author} {\bibfnamefont {J.~T.}\ \bibnamefont
  {Liu}},\ }\href {\doibase 10.1007/JHEP07(2020)073} {\bibfield  {journal}
  {\bibinfo  {journal} {JHEP}\ }\textbf {\bibinfo {volume} {07}},\ \bibinfo
  {pages} {073} (\bibinfo {year} {2020})},\ \Eprint
  {http://arxiv.org/abs/1912.04169} {arXiv:1912.04169 [hep-th]} \BibitemShut
  {NoStop}%
\bibitem [{\citenamefont {Kinney}\ \emph {et~al.}(2007)\citenamefont {Kinney},
  \citenamefont {Maldacena}, \citenamefont {Minwalla},\ and\ \citenamefont
  {Raju}}]{Kinney:2005ej}%
  \BibitemOpen
  \bibfield  {author} {\bibinfo {author} {\bibfnamefont {J.}~\bibnamefont
  {Kinney}}, \bibinfo {author} {\bibfnamefont {J.~M.}\ \bibnamefont
  {Maldacena}}, \bibinfo {author} {\bibfnamefont {S.}~\bibnamefont {Minwalla}},
  \ and\ \bibinfo {author} {\bibfnamefont {S.}~\bibnamefont {Raju}},\ }\href
  {\doibase 10.1007/s00220-007-0258-7} {\bibfield  {journal} {\bibinfo
  {journal} {Commun. Math. Phys.}\ }\textbf {\bibinfo {volume} {275}},\
  \bibinfo {pages} {209} (\bibinfo {year} {2007})},\ \Eprint
  {http://arxiv.org/abs/hep-th/0510251} {arXiv:hep-th/0510251} \BibitemShut
  {NoStop}%
\bibitem [{\citenamefont {Rastelli}\ and\ \citenamefont
  {Razamat}(2017)}]{Rastelli:2016tbz}%
  \BibitemOpen
  \bibfield  {author} {\bibinfo {author} {\bibfnamefont {L.}~\bibnamefont
  {Rastelli}}\ and\ \bibinfo {author} {\bibfnamefont {S.~S.}\ \bibnamefont
  {Razamat}},\ }\href {\doibase 10.1088/1751-8121/aa76a6} {\bibfield  {journal}
  {\bibinfo  {journal} {J. Phys. A}\ }\textbf {\bibinfo {volume} {50}},\
  \bibinfo {pages} {443013} (\bibinfo {year} {2017})},\ \Eprint
  {http://arxiv.org/abs/1608.02965} {arXiv:1608.02965 [hep-th]} \BibitemShut
  {NoStop}%
\bibitem [{\citenamefont {Gadde}(2020)}]{Gadde:2020yah}%
  \BibitemOpen
  \bibfield  {author} {\bibinfo {author} {\bibfnamefont {A.}~\bibnamefont
  {Gadde}},\ }\href@noop {} {\  (\bibinfo {year} {2020})},\ \Eprint
  {http://arxiv.org/abs/2006.13630} {arXiv:2006.13630 [hep-th]} \BibitemShut
  {NoStop}%
\bibitem [{\citenamefont {Hosseini}\ \emph {et~al.}(2017)\citenamefont
  {Hosseini}, \citenamefont {Hristov},\ and\ \citenamefont
  {Zaffaroni}}]{Hosseini:2017mds}%
  \BibitemOpen
  \bibfield  {author} {\bibinfo {author} {\bibfnamefont {S.~M.}\ \bibnamefont
  {Hosseini}}, \bibinfo {author} {\bibfnamefont {K.}~\bibnamefont {Hristov}}, \
  and\ \bibinfo {author} {\bibfnamefont {A.}~\bibnamefont {Zaffaroni}},\ }\href
  {\doibase 10.1007/JHEP07(2017)106} {\bibfield  {journal} {\bibinfo  {journal}
  {JHEP}\ }\textbf {\bibinfo {volume} {07}},\ \bibinfo {pages} {106} (\bibinfo
  {year} {2017})},\ \Eprint {http://arxiv.org/abs/1705.05383} {arXiv:1705.05383
  [hep-th]} \BibitemShut {NoStop}%
\bibitem [{\citenamefont {Cabo-Bizet}\ \emph
  {et~al.}(2019{\natexlab{a}})\citenamefont {Cabo-Bizet}, \citenamefont
  {Cassani}, \citenamefont {Martelli},\ and\ \citenamefont
  {Murthy}}]{Cabo-Bizet:2018ehj}%
  \BibitemOpen
  \bibfield  {author} {\bibinfo {author} {\bibfnamefont {A.}~\bibnamefont
  {Cabo-Bizet}}, \bibinfo {author} {\bibfnamefont {D.}~\bibnamefont {Cassani}},
  \bibinfo {author} {\bibfnamefont {D.}~\bibnamefont {Martelli}}, \ and\
  \bibinfo {author} {\bibfnamefont {S.}~\bibnamefont {Murthy}},\ }\href
  {\doibase 10.1007/JHEP10(2019)062} {\bibfield  {journal} {\bibinfo  {journal}
  {JHEP}\ }\textbf {\bibinfo {volume} {10}},\ \bibinfo {pages} {062} (\bibinfo
  {year} {2019}{\natexlab{a}})},\ \Eprint {http://arxiv.org/abs/1810.11442}
  {arXiv:1810.11442 [hep-th]} \BibitemShut {NoStop}%
\bibitem [{\citenamefont {Choi}\ \emph
  {et~al.}(2018{\natexlab{a}})\citenamefont {Choi}, \citenamefont {Kim},
  \citenamefont {Kim},\ and\ \citenamefont {Nahmgoong}}]{Choi:2018hmj}%
  \BibitemOpen
  \bibfield  {author} {\bibinfo {author} {\bibfnamefont {S.}~\bibnamefont
  {Choi}}, \bibinfo {author} {\bibfnamefont {J.}~\bibnamefont {Kim}}, \bibinfo
  {author} {\bibfnamefont {S.}~\bibnamefont {Kim}}, \ and\ \bibinfo {author}
  {\bibfnamefont {J.}~\bibnamefont {Nahmgoong}},\ }\href@noop {} {\  (\bibinfo
  {year} {2018}{\natexlab{a}})},\ \Eprint {http://arxiv.org/abs/1810.12067}
  {arXiv:1810.12067 [hep-th]} \BibitemShut {NoStop}%
\bibitem [{\citenamefont {Choi}\ \emph
  {et~al.}(2018{\natexlab{b}})\citenamefont {Choi}, \citenamefont {Kim},
  \citenamefont {Kim},\ and\ \citenamefont {Nahmgoong}}]{Choi:2018vbz}%
  \BibitemOpen
  \bibfield  {author} {\bibinfo {author} {\bibfnamefont {S.}~\bibnamefont
  {Choi}}, \bibinfo {author} {\bibfnamefont {J.}~\bibnamefont {Kim}}, \bibinfo
  {author} {\bibfnamefont {S.}~\bibnamefont {Kim}}, \ and\ \bibinfo {author}
  {\bibfnamefont {J.}~\bibnamefont {Nahmgoong}},\ }\href@noop {} {\  (\bibinfo
  {year} {2018}{\natexlab{b}})},\ \Eprint {http://arxiv.org/abs/1811.08646}
  {arXiv:1811.08646 [hep-th]} \BibitemShut {NoStop}%
\bibitem [{\citenamefont {Closset}\ \emph {et~al.}(2017)\citenamefont
  {Closset}, \citenamefont {Kim},\ and\ \citenamefont
  {Willett}}]{Closset:2017zgf}%
  \BibitemOpen
  \bibfield  {author} {\bibinfo {author} {\bibfnamefont {C.}~\bibnamefont
  {Closset}}, \bibinfo {author} {\bibfnamefont {H.}~\bibnamefont {Kim}}, \ and\
  \bibinfo {author} {\bibfnamefont {B.}~\bibnamefont {Willett}},\ }\href
  {\doibase 10.1007/JHEP03(2017)074} {\bibfield  {journal} {\bibinfo  {journal}
  {JHEP}\ }\textbf {\bibinfo {volume} {03}},\ \bibinfo {pages} {074} (\bibinfo
  {year} {2017})},\ \Eprint {http://arxiv.org/abs/1701.03171} {arXiv:1701.03171
  [hep-th]} \BibitemShut {NoStop}%
\bibitem [{\citenamefont {Benini}\ and\ \citenamefont
  {Milan}(2020{\natexlab{b}})}]{Benini:2018mlo}%
  \BibitemOpen
  \bibfield  {author} {\bibinfo {author} {\bibfnamefont {F.}~\bibnamefont
  {Benini}}\ and\ \bibinfo {author} {\bibfnamefont {P.}~\bibnamefont {Milan}},\
  }\href {\doibase 10.1007/s00220-019-03679-y} {\bibfield  {journal} {\bibinfo
  {journal} {Commun. Math. Phys.}\ }\textbf {\bibinfo {volume} {376}},\
  \bibinfo {pages} {1413} (\bibinfo {year} {2020}{\natexlab{b}})},\ \Eprint
  {http://arxiv.org/abs/1811.04107} {arXiv:1811.04107 [hep-th]} \BibitemShut
  {NoStop}%
\bibitem [{\citenamefont {Gonz\'alez~Lezcano}\ and\ \citenamefont
  {Pando~Zayas}(2020)}]{Lezcano:2019pae}%
  \BibitemOpen
  \bibfield  {author} {\bibinfo {author} {\bibfnamefont {A.}~\bibnamefont
  {Gonz\'alez~Lezcano}}\ and\ \bibinfo {author} {\bibfnamefont {L.~A.}\
  \bibnamefont {Pando~Zayas}},\ }\href {\doibase 10.1007/JHEP03(2020)088}
  {\bibfield  {journal} {\bibinfo  {journal} {JHEP}\ }\textbf {\bibinfo
  {volume} {03}},\ \bibinfo {pages} {088} (\bibinfo {year} {2020})},\ \Eprint
  {http://arxiv.org/abs/1907.12841} {arXiv:1907.12841 [hep-th]} \BibitemShut
  {NoStop}%
\bibitem [{\citenamefont {Lanir}\ \emph {et~al.}(2020)\citenamefont {Lanir},
  \citenamefont {Nedelin},\ and\ \citenamefont {Sela}}]{Lanir:2019abx}%
  \BibitemOpen
  \bibfield  {author} {\bibinfo {author} {\bibfnamefont {A.}~\bibnamefont
  {Lanir}}, \bibinfo {author} {\bibfnamefont {A.}~\bibnamefont {Nedelin}}, \
  and\ \bibinfo {author} {\bibfnamefont {O.}~\bibnamefont {Sela}},\ }\href
  {\doibase 10.1007/JHEP04(2020)091} {\bibfield  {journal} {\bibinfo  {journal}
  {JHEP}\ }\textbf {\bibinfo {volume} {04}},\ \bibinfo {pages} {091} (\bibinfo
  {year} {2020})},\ \Eprint {http://arxiv.org/abs/1908.01737} {arXiv:1908.01737
  [hep-th]} \BibitemShut {NoStop}%
\bibitem [{\citenamefont {Honda}(2019)}]{Honda:2019cio}%
  \BibitemOpen
  \bibfield  {author} {\bibinfo {author} {\bibfnamefont {M.}~\bibnamefont
  {Honda}},\ }\href {\doibase 10.1103/PhysRevD.100.026008} {\bibfield
  {journal} {\bibinfo  {journal} {Phys. Rev. D}\ }\textbf {\bibinfo {volume}
  {100}},\ \bibinfo {pages} {026008} (\bibinfo {year} {2019})},\ \Eprint
  {http://arxiv.org/abs/1901.08091} {arXiv:1901.08091 [hep-th]} \BibitemShut
  {NoStop}%
\bibitem [{\citenamefont {Arabi~Ardehali}(2019)}]{ArabiArdehali:2019tdm}%
  \BibitemOpen
  \bibfield  {author} {\bibinfo {author} {\bibfnamefont {A.}~\bibnamefont
  {Arabi~Ardehali}},\ }\href {\doibase 10.1007/JHEP06(2019)134} {\bibfield
  {journal} {\bibinfo  {journal} {JHEP}\ }\textbf {\bibinfo {volume} {06}},\
  \bibinfo {pages} {134} (\bibinfo {year} {2019})},\ \Eprint
  {http://arxiv.org/abs/1902.06619} {arXiv:1902.06619 [hep-th]} \BibitemShut
  {NoStop}%
\bibitem [{\citenamefont {Cabo-Bizet}\ \emph
  {et~al.}(2019{\natexlab{b}})\citenamefont {Cabo-Bizet}, \citenamefont
  {Cassani}, \citenamefont {Martelli},\ and\ \citenamefont
  {Murthy}}]{Cabo-Bizet:2019osg}%
  \BibitemOpen
  \bibfield  {author} {\bibinfo {author} {\bibfnamefont {A.}~\bibnamefont
  {Cabo-Bizet}}, \bibinfo {author} {\bibfnamefont {D.}~\bibnamefont {Cassani}},
  \bibinfo {author} {\bibfnamefont {D.}~\bibnamefont {Martelli}}, \ and\
  \bibinfo {author} {\bibfnamefont {S.}~\bibnamefont {Murthy}},\ }\href
  {\doibase 10.1007/JHEP08(2019)120} {\bibfield  {journal} {\bibinfo  {journal}
  {JHEP}\ }\textbf {\bibinfo {volume} {08}},\ \bibinfo {pages} {120} (\bibinfo
  {year} {2019}{\natexlab{b}})},\ \Eprint {http://arxiv.org/abs/1904.05865}
  {arXiv:1904.05865 [hep-th]} \BibitemShut {NoStop}%
\bibitem [{\citenamefont {Amariti}\ \emph {et~al.}(2019)\citenamefont
  {Amariti}, \citenamefont {Garozzo},\ and\ \citenamefont
  {Lo~Monaco}}]{Amariti:2019mgp}%
  \BibitemOpen
  \bibfield  {author} {\bibinfo {author} {\bibfnamefont {A.}~\bibnamefont
  {Amariti}}, \bibinfo {author} {\bibfnamefont {I.}~\bibnamefont {Garozzo}}, \
  and\ \bibinfo {author} {\bibfnamefont {G.}~\bibnamefont {Lo~Monaco}},\
  }\href@noop {} {\  (\bibinfo {year} {2019})},\ \Eprint
  {http://arxiv.org/abs/1904.10009} {arXiv:1904.10009 [hep-th]} \BibitemShut
  {NoStop}%
\bibitem [{\citenamefont {Cabo-Bizet}\ and\ \citenamefont
  {Murthy}(2020)}]{Cabo-Bizet:2019eaf}%
  \BibitemOpen
  \bibfield  {author} {\bibinfo {author} {\bibfnamefont {A.}~\bibnamefont
  {Cabo-Bizet}}\ and\ \bibinfo {author} {\bibfnamefont {S.}~\bibnamefont
  {Murthy}},\ }\href {\doibase 10.1007/JHEP09(2020)184} {\bibfield  {journal}
  {\bibinfo  {journal} {JHEP}\ }\textbf {\bibinfo {volume} {09}},\ \bibinfo
  {pages} {184} (\bibinfo {year} {2020})},\ \Eprint
  {http://arxiv.org/abs/1909.09597} {arXiv:1909.09597 [hep-th]} \BibitemShut
  {NoStop}%
\bibitem [{\citenamefont {Cabo-Bizet}\ \emph {et~al.}(2020)\citenamefont
  {Cabo-Bizet}, \citenamefont {Cassani}, \citenamefont {Martelli},\ and\
  \citenamefont {Murthy}}]{Cabo-Bizet:2020nkr}%
  \BibitemOpen
  \bibfield  {author} {\bibinfo {author} {\bibfnamefont {A.}~\bibnamefont
  {Cabo-Bizet}}, \bibinfo {author} {\bibfnamefont {D.}~\bibnamefont {Cassani}},
  \bibinfo {author} {\bibfnamefont {D.}~\bibnamefont {Martelli}}, \ and\
  \bibinfo {author} {\bibfnamefont {S.}~\bibnamefont {Murthy}},\ }\href
  {\doibase 10.1007/JHEP11(2020)150} {\bibfield  {journal} {\bibinfo  {journal}
  {JHEP}\ }\textbf {\bibinfo {volume} {11}},\ \bibinfo {pages} {150} (\bibinfo
  {year} {2020})},\ \Eprint {http://arxiv.org/abs/2005.10654} {arXiv:2005.10654
  [hep-th]} \BibitemShut {NoStop}%
\bibitem [{\citenamefont {Agarwal}\ \emph {et~al.}(2020)\citenamefont
  {Agarwal}, \citenamefont {Choi}, \citenamefont {Kim}, \citenamefont {Kim},\
  and\ \citenamefont {Nahmgoong}}]{Agarwal:2020zwm}%
  \BibitemOpen
  \bibfield  {author} {\bibinfo {author} {\bibfnamefont {P.}~\bibnamefont
  {Agarwal}}, \bibinfo {author} {\bibfnamefont {S.}~\bibnamefont {Choi}},
  \bibinfo {author} {\bibfnamefont {J.}~\bibnamefont {Kim}}, \bibinfo {author}
  {\bibfnamefont {S.}~\bibnamefont {Kim}}, \ and\ \bibinfo {author}
  {\bibfnamefont {J.}~\bibnamefont {Nahmgoong}},\ }\href@noop {} {\  (\bibinfo
  {year} {2020})},\ \Eprint {http://arxiv.org/abs/2005.11240} {arXiv:2005.11240
  [hep-th]} \BibitemShut {NoStop}%
\bibitem [{\citenamefont {Benini}\ \emph {et~al.}(2020)\citenamefont {Benini},
  \citenamefont {Colombo}, \citenamefont {Soltani}, \citenamefont {Zaffaroni},\
  and\ \citenamefont {Zhang}}]{Benini:2020gjh}%
  \BibitemOpen
  \bibfield  {author} {\bibinfo {author} {\bibfnamefont {F.}~\bibnamefont
  {Benini}}, \bibinfo {author} {\bibfnamefont {E.}~\bibnamefont {Colombo}},
  \bibinfo {author} {\bibfnamefont {S.}~\bibnamefont {Soltani}}, \bibinfo
  {author} {\bibfnamefont {A.}~\bibnamefont {Zaffaroni}}, \ and\ \bibinfo
  {author} {\bibfnamefont {Z.}~\bibnamefont {Zhang}},\ }\href {\doibase
  10.1088/1361-6382/abb39b} {\bibfield  {journal} {\bibinfo  {journal} {Class.
  Quant. Grav.}\ }\textbf {\bibinfo {volume} {37}},\ \bibinfo {pages} {215021}
  (\bibinfo {year} {2020})},\ \Eprint {http://arxiv.org/abs/2005.12308}
  {arXiv:2005.12308 [hep-th]} \BibitemShut {NoStop}%
\bibitem [{\citenamefont {Gonz\'alez~Lezcano}\ \emph
  {et~al.}(2020)\citenamefont {Gonz\'alez~Lezcano}, \citenamefont {Hong},
  \citenamefont {Liu},\ and\ \citenamefont
  {Pando~Zayas}}]{GonzalezLezcano:2020yeb}%
  \BibitemOpen
  \bibfield  {author} {\bibinfo {author} {\bibfnamefont {A.}~\bibnamefont
  {Gonz\'alez~Lezcano}}, \bibinfo {author} {\bibfnamefont {J.}~\bibnamefont
  {Hong}}, \bibinfo {author} {\bibfnamefont {J.~T.}\ \bibnamefont {Liu}}, \
  and\ \bibinfo {author} {\bibfnamefont {L.~A.}\ \bibnamefont {Pando~Zayas}},\
  }\href@noop {} {\  (\bibinfo {year} {2020})},\ \Eprint
  {http://arxiv.org/abs/2007.12604} {arXiv:2007.12604 [hep-th]} \BibitemShut
  {NoStop}%
\bibitem [{\citenamefont {Zaffaroni}(2019)}]{Zaffaroni:2019dhb}%
  \BibitemOpen
  \bibfield  {author} {\bibinfo {author} {\bibfnamefont {A.}~\bibnamefont
  {Zaffaroni}}\ }(\bibinfo {year} {2019})\ \Eprint
  {http://arxiv.org/abs/1902.07176} {arXiv:1902.07176 [hep-th]} \BibitemShut
  {NoStop}%
\bibitem [{\citenamefont {Goldstein}\ \emph {et~al.}(2020)\citenamefont
  {Goldstein}, \citenamefont {Jejjala}, \citenamefont {Lei}, \citenamefont {van
  Leuven},\ and\ \citenamefont {Li}}]{Goldstein:2020yvj}%
  \BibitemOpen
  \bibfield  {author} {\bibinfo {author} {\bibfnamefont {K.}~\bibnamefont
  {Goldstein}}, \bibinfo {author} {\bibfnamefont {V.}~\bibnamefont {Jejjala}},
  \bibinfo {author} {\bibfnamefont {Y.}~\bibnamefont {Lei}}, \bibinfo {author}
  {\bibfnamefont {S.}~\bibnamefont {van Leuven}}, \ and\ \bibinfo {author}
  {\bibfnamefont {W.}~\bibnamefont {Li}},\ }\href@noop {} {\  (\bibinfo {year}
  {2020})},\ \Eprint {http://arxiv.org/abs/2011.06605} {arXiv:2011.06605
  [hep-th]} \BibitemShut {NoStop}%
\bibitem [{\citenamefont {Cabo-Bizet}(2020)}]{Cabo-Bizet:2020ewf}%
  \BibitemOpen
  \bibfield  {author} {\bibinfo {author} {\bibfnamefont {A.}~\bibnamefont
  {Cabo-Bizet}},\ }\href@noop {} {\  (\bibinfo {year} {2020})},\ \Eprint
  {http://arxiv.org/abs/2012.04815} {arXiv:2012.04815 [hep-th]} \BibitemShut
  {NoStop}%
\end{thebibliography}%
\end{document}